\documentclass[10pt, oneside, a4paper, twocolumn]{article}

 \usepackage{rotating}
\usepackage[cjk]{kotex}
\usepackage{setspace}
\usepackage{multirow}
\usepackage[a4paper,left=20mm,right=20mm,top=30mm,bottom=30mm]{geometry}

\usepackage[letterspace=200]{microtype}
\usepackage{tikz}
\usepackage{graphicx}
\usepackage{amssymb}
\usepackage{endnotes}
\usepackage{amsmath}
\usepackage[normalem]{ulem}
\usepackage{booktabs}
\usepackage{tabularx}
\usepackage{array}
\usepackage{fancyhdr}
\useunder{\uline}{\ul}{}
\usepackage{amsthm}
\graphicspath{ {fig/} }

\newcolumntype{Y}{>{\centering\arraybackslash}X}

\begin{document}

\onecolumn
\setstretch{1.2}

\title{Information Content of Financial Youtube Channel: Case Study of 3PROTV and Korean Stock Market}

\author{HyeonJun Kim$^{*}$\medskip\\{\normalsize School of Finance, Soongsil University}
\medskip\\{\normalsize$^*$inertia164214@soongsil.ac.kr or hyeonjunacademic@gmail.com}}


\maketitle

\begin{abstract}

We investigate the information content of 3PROTV, a south Korean financial youtube channel. In our sample we found evidence for the hypothesis that the channel have information content on stock selection, but only on negative sentiment. Positively mentioned stock had pre-announcement spike followed by steep fall in stock price around announcement period. Negatively mentioned stock started underperforming around the announcement period, with underreaction dynamics in post-announcement period. In the area of market timing, we found that change of sentimental tone of 3PROTV than its historical average predicts the lead value of Korean market portfolio return. Its predictive power cannot be explained by future change in news sentiment, future short term interest rate, and future liquidity risk. \\\\

\textbf{keywords:} {[Information Content], [YouTube], [Market Timing], [Stock Selection]}


\end{abstract}

\setstretch{1.5}

\section{Introduction}

An information material have information content, in Beaver(1968)'s term, "if it leads to a change in investors' assessments of the probability distribution of future returns (or prices), such that there is a change in equilibrium value of the current market price". \cite{beaver1968information} This argument implies that information content causes future price moves and could be used for future return prediction. There are trivial source of information contents like earning reports or president's letters to shareholders information. \cite{abrahamson1996information,beaver1968information} 

In that sense, it will be somewhat abnormal for a normal YouTube content creator to have an information of such impact. There are some studies that investigates the relationship between social media reaction and stock price, and few interpreted that it have information content. \cite{sul2014trading} However it is questionable that this effects are because of whether mispricing caused by sentiment, or change in permanent equilibrium value. Though it could be agreed that its economical significance cannot be ignored. \cite{sul2017trading}. 

For more robust theoretical foundation, we can consider more satisfying source, financial experts. For example, there is evidence that analyst reports have information content even controlling for confounding firm events. \cite{asquith2005information} 
Our subject in interest is at the mix of two extremes, social media content and professional content. The platform of the content is social media, therefore the influence of noise traders is inevitable. However, the content creators are financial experts, which opens up the possibility of creating new information by researching for undiscovered data or interpreting existing data. 

Here we focus on 3PROTV, one of the example of expert-run YouTube channel. 3PROTV, including its YouTube channel is expected to worth almost 200 million dollars to date, and it has been one of the most rapidly grown Financial YouTube channels in South Korea. It's popularity had significantly grown during COVID-19 pandemic period, when quantitative easing policies cause the stock market to spike, which attracted new investors and demands for investment information had risen. The impact of the channel's content do seems socially significant, however is unclear that it will have information about the future stock market. In this paper, We investigate the latter. 

The result of the analysis is quite unexpected. Roughly speaking, 3PROTV's content does have information content in our sample period, both on stock selection and market timing. Stocks that are negatively mentioned by 3PROTV had underperform after the announcement, while positively mentioned stock had outperformed before the announcement. If we consider 3PROTV's information as sentimental, the result will be inconsistent with the prediction that current literature like Baker and Wurgler (2007) suggest, which predicts negative sentiment causes future outperformance, and vice versa. \cite{baker2007investor} Another surprising part is that 3PROTV's content sentiment, when preprocessed properly, can serve as future return prediction indicator. We developed a sentiment index using 3PROTV's content sentiment and found robust relation between future return and the index regradless of index parameter. The daily trading strategy using our sentiment index show outperformance from 11.5\% to 23.6\% without transaction cost. The predictability persisted even controlling for future news sentiment, short term interest rate, and liquidity risk factor ($PctZero$), making 3PROTV's content information somewhat independent from those information. 
In section 2 we show our methodology, and in section 3 we present our empirical results. In section 4 we conclude the study with some remarks.

\section{Methodology}

For analyzing the 3PROTV's specific content, we scraped YouTube's auto-transcription data to extract the mentioned stock and script's overall sentiment through August 2, 2022 to November 28, 2022, which in total a 500 content scripts. First, we concatenate the script within the same day and tokenize, i.e. decompose the each content's script such that the script becomes a list of phased words. We then match the tokenized words of the script with DataGuide's Korean firm name list. We exclude some common noun-like stock names like "태양"(sun in Korean), "대상"(subject, grand prize, etc. in Korean), which might be mentioned because of other reasons. Another attempts to lower the selection bias is to only select the stock name into our sample only when the stock name is mentioned more than 2 times. By this method, we ensure that those stocks represents the main theme of that day's script, while lowering the chance of a stock with low relevance to the content being accidentally included in the sample.

For sentiment analysis, we use KOSELF(Korean Sentiment Lexicon for Finance) \cite{cho2021building} to extract the sentiment level of each 3PROTV's content script. KOSELF is a Korean Sentiment Lexicon exclusively developed for financial usage, meaning that it will detect financial setiment in scripts better than other alternatives like KOSAC or KNU's sentiment lexicon, which are developed for more general sentiment analysis.
We use the sentiment level of content release date $i$ $SENTIMENT^{(i)}$ , which is calculated using formula \ref{sentiment}.
\begin{equation} \label{sentiment}
SENTIMENT^{(i)}={N_{positive}^{(i)} - N_{negative}^{(i)} \over N_{positive}^{(i)} +N_{negative}^{(i)}}
\end{equation}
Where $N_{positive}^{(i)}$ is the number of words which is in content script at date $i$ and also a KOSELF positive word, and $N_{negative}^{(i)}$ is the number of words which is in content script at date $i$ and also a KOSELF negative word. We count duplicate words separately.
For separating the positively mentioned stocks and negatively mentioned stocks, we match $SENTIMENT^{(i)}$ to each stock $j$ that is mentioned in content script at date $i$.

Our apporach for evaluating the information content of the 3PROTV scripts are in two main ways. One is analyzing its stock selection information, and the other is its market timing. 
For investigating the possibility of information content on stock selection, we deploy the standard event study methodology, \cite{peterson1989event} but with some specific settings. One of those differences is that we use state-of-the-art expected return model, the Fama-French 5 factor model, which takes a form like formula. \cite{fama2015five} 
\begin{equation}
    R_j(t)-r_f(t)=\alpha_j + \beta_j \cdot RMRF(t) + s_j \cdot SMB(t) + h_j \cdot HML(t) + r_j \cdot RMW(t) + c_j \cdot CMA(t) + \epsilon_j (t)
\end{equation}
Where $R_j(t)$ is the daily investment return of stock $j$ at day  $t$, $r_f(t)$ is the risk-free asset's daily investment return at day $t$, $RMRF(t)$ is the market portfolio's daily excess return at day $t$, $SMB(t)$ is a daily return of zero-investment portfolio (or a long-short portfolio) of small firm portfolio minus big firm portfolio at day $t$, $HML(t)$ is a daily return of zero-investment portfolio of high value (high book-to-market ratio) firm portfolio minus low value  (low book-to-market ratio) firm portfolio at day $t$, $RMW(t)$ is a daily return of zero-investment portfolio of robust (high operating income-book value ratio) firm portfolio minus weak (low operating income-book value ratio) firm portfolio at day $t$, $CMA(t)$ is a daily return of zero-investment portfolio of conservative (low asset growth) firm portfolio minus aggressive (high asset growth) firm portfolio at day $t$, and $\epsilon_j(t)$ is a idiosyncratic variation of stock return. We followed Fama and French (2015)'s method as same as possible while applying it on the Korean stock market data gathered from DataGuide, with some adaption like using KOSPI breakpoints to sort the stocks in KOSPI and KOSDAQ, which correspond to using NYSE breakpoints in the original paper.
Also, we extend the model to have 1 leading and 1 lagging variables for each factors to consider delayed or advance reaction, totaling up to 15 independent variables for the expected return estimation.

Our event study methodology is as follows. First, based on relative date $0$ when the stock $j$ is mentioned by 3PROTV's panel, we set a estimation period from $-273$ to $-21$ for estimating Fama-French 5 factor model's exposures (i.e. the slopes like $\beta_j, s_j, \cdots$) and intercept $\alpha_j$ parameters for each stock. Then for $-20$ to $20$, the expected daily return $\overline{R_j}(t)$ of the stock is calculated. We then derive abnormal return $AR_j(t)$, which is nothing but
\begin{equation}
    AR_j(t) = R_j(t)-\overline{R_j}(t)
\end{equation}
We then average the $AR_j(t)$ by stocks by certain list of stocks $J$ to obtain the average abnormal return $AAR_j(t)$, i.e.
\begin{equation}
    AAR_j(t) = {1\over N}\sum_{j\in J} AR_j(t)
\end{equation}
where $N$ is the total number of stocks in $J$. Finally, we sum the $AAR_j(t)$ over the $T$ days in the event window or sub-window $[t_0, t_0+T]$ to form the cumulative average abnormal return $CAAR_j(t_0+T)$
\begin{equation}
    CAAR_j(t_0, t_0+T) = \sum_{t=t_0}^{t_0+T} AAR_j(t)
\end{equation}
We set $t_0=-20$ and $T=40$ for full event window, with additional analysis using $t_0=0$ and $T=20$ for event window period where the investor is notified by 3PROTV. Then we compare pattern of CAAR between $J^{(+)}$ and  $J^{(-)}$, where $J^{(+)}$ consist of stock with corresponding $SENTIMENT^{(i)}$ being positive, and $J^{(-)}$  being negative. By this rough analysis, we get a sense of average stock price change patterns of which 3PROTV had flagged positive or negative. Comparing the two CAARs time series will show whether 3PROTV's information is already public (whether or not there is a price reaction before the mention), or the accuracy of 3PROTV's sentiment prediction (positively flagged stock should outperform negatively flagged stocks). Sample size of postively mentioned stocks are total 13, while negatively mentioned stocks are total 54. The asymmetry is expected, due to the general negative sentiment of 2022's stock market investors.

For showing evidence for market timing information content, we develop a preprocessed sentiment index $S(N, t)$ from $SENTIMENT^{(i)}$, which we demonstrate that it has predictive power for sign of 2day leading market portfolio return. 
We define a sentiment index $S(N, t)$ with a binary value that is defined as equation \ref{s}.
\begin{equation}\label{s}
S(N, t) =\begin{cases}
 1 & \text{if  } SENTIMENT^{(t)} - {1\over N}\sum_{i=t-N-1}^{t} SENTIMENT^{(i)} > 0 \\
0 & \text{else} \\
\end{cases}
\end{equation}

We report $R^2$ between $S(N, t-2)$ and $RMRF(t)$ with several different $N$s for robustness check. Then we check the profitability of market timing using $S(N, t-2)$.
Lastly we consider the logit regression with $S(N, t-2)$ as its independent variable and the sign of $RMRF(t)$ as its dependent variable. We then control for change in news sentiment index (NSI) \cite{seo2022machine} whose daily values are announced by the Bank of Korea, with also business cycle variable like short interest rate and liquidity risk factor ($PctZero$). Interest rate for 91day Monetary Stabilization Bond is used as short interest rate and $PctZero$ is a daily ratio of zero-return stock within the market. Controlling for business cycle variable is motivated by Sibley, Wang, Xing and Zhang (2016), who suggested that baker-Wurgler sentiment index have cross-sectional return prediction ability mainly due to the information content related to business cycle variable. \cite{sibley2016information}

\section{Empirical Results}

\subsection{Information Content on Stock Selection}

\begin{figure}[h!]\label{ccar1}
\caption{Time series plot of cumulative average abnormal return of two distinct stock group, the blue curve being a negatively flagged stock by 3PROTV, and orange curve being a positivly flagged stock by 3PROTV. The vertical line is the announcement (mention) time, which is relatively zero. If the announcement was not a trading day, we move the announcement date to the nearlest future trading day.}
\centering
\includegraphics[width=0.75\textwidth]{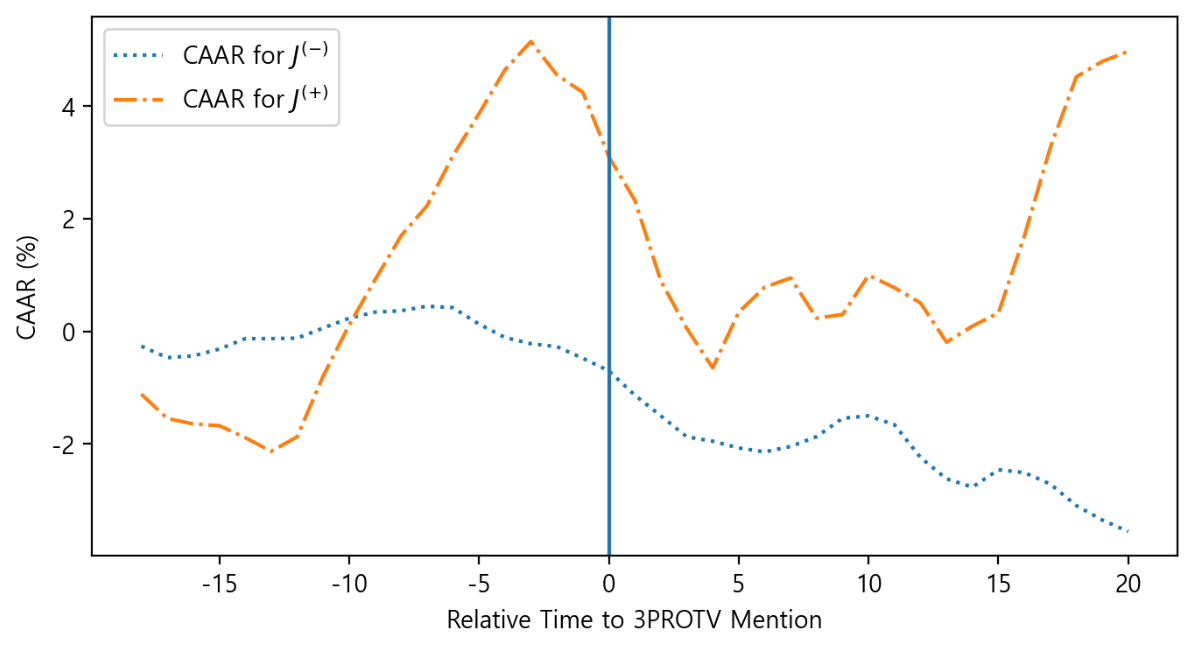}
\end{figure}

\begin{figure}[h!]\label{ccar11}
\caption{Time series plot of cumulative average abnormal return of two distinct stock group, the blue curve being a negatively flagged stock by 3PROTV, and orange curve being a positivly flagged stock by 3PROTV. We limit the event window The vertical line is the announcement (mention) time, which is relatively zero. If the announcement was not a trading day, we move the announcement date to the nearlest future trading day.}
\centering
\includegraphics[width=0.75\textwidth]{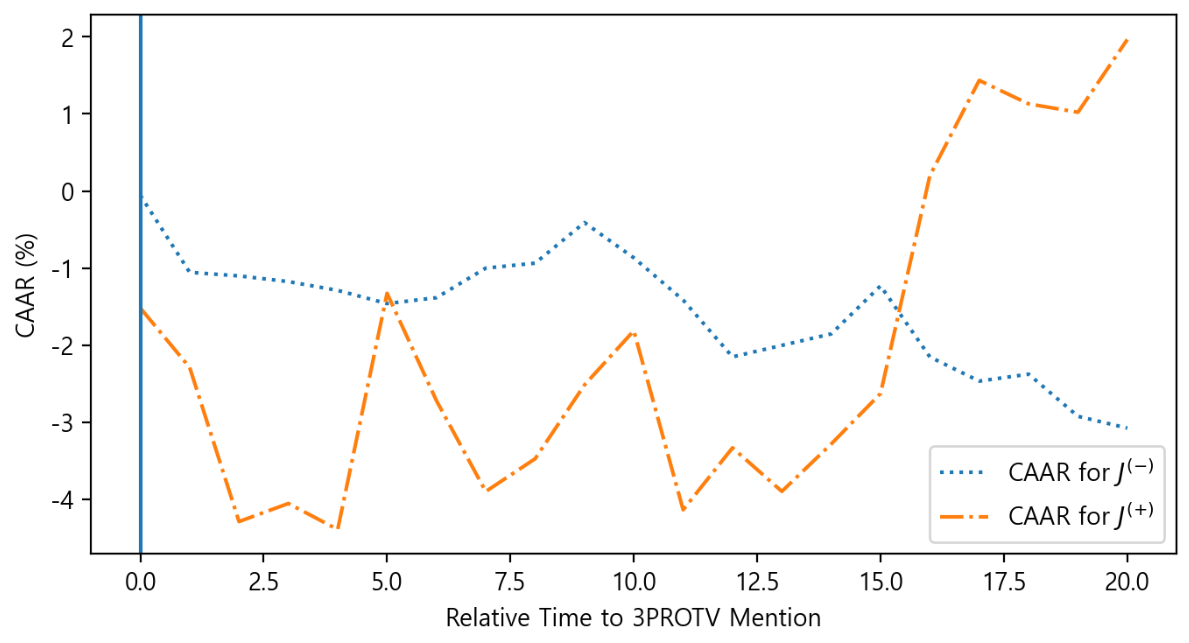}
\end{figure}

Figure \ref{ccar1} and Figure \ref{ccar11} show the result of the first analysis using CCAR. We can found that stocks in $J^{(+)}$ outperforms in average than stocks in $J^{(-)}$ within a event window, which means 3PROTV's content sentiment roughly have information content on the average stock performance. However comparing the two figures, it seems that positive content sentiment information is already being incorporated into the stock price before 12 days, with spike in stock price right before the announcement. This implies that 3PROTV's positively flagged stocks and their information are likely to be already known to the public. 

Information with negative sentiment seems to have more new information content, which price shock happening around the announcement date. The lingering shock throughout the post-announcement period resembles the investor's  disposition effect\cite{grinblatt2002disposition}, as negative price shock discourages investors to sell their stocks, forming an underreaction.

Long story short, 3PROTV's content sentiment on stocks have independent information content only on stock selection information with negative sentiment. Information with positive sentiment have little information content, with limitation to investment decision. Investors in these stocks faces overreacted and volatile stock price.

\subsection{Information Content on Market Timing}

Considering the former result, it could be suspected that the script sentiment can be used in other application. We focus on short-term market return prediction.

\begin{figure}[h!]\label{r2}
\caption{$R^2$ between $S(N, t-2)$ and $RMRF(t)$ with difference. The dotted line is $R^2 = 1.5\%$, which is the baseline performance for usable signal. $R^2$ is the square of correlation between $S(N, t-2)$ and $RMRF(t)$. We report that for every $N$, the correlation is positive.}
\centering
\includegraphics[width=0.75\textwidth]{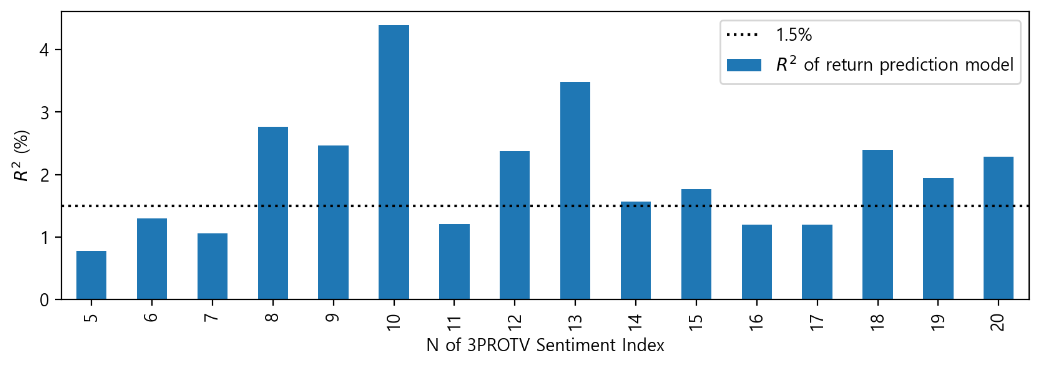}
\end{figure}

We had mentioned the formula for $S(N, t)$, which is one when the current sentiment of 3PROTV's script is higher than its N moving average, and zero if it is not. This methodology maximize the predictability in the sample, as it is shown in Figure \ref{r2} . from $N$ ranging from 5 to 20, 10 out of 16 index at $t-2$ show more than 1.5\% of $R^2$, which is quite a high bar for market portfolio return prediction model. We choose $t-2$ because the youtube content could be released after the market is closed, making it impossible to trade stocks using index at $t-1$. Some might argue that the performance of $S(N, t)$ is due to data snooping, not a real discovery. While it is true that $S(N, t)$ is not tested for out-of-sample data (which will be left for follow-up study), we dispute that argument with some key point. One of two points is that $S(N, t)$ have less information than $SENTIMENT^{(i)}$, which works as a penalty rather than advantage. The other point is that at least the index is robust to change in $N$, which the deployment of moving average is less data mining than it seems. All things aside, it is reasonable to have a healthy scepticism about our methodology, and replication study by using data after 2022 is required for further application.

We apply the sentiment index to trading using the "Sentiment Strategy", where the trader long $RMRF(t)$ when $S(N, t-2)$ is 1 and short when $S(N, t-2)$ is 0. We tested with setting the parameters to $N=5, 10, 20$, which in return show outperformance ranging from 11.5\% (N=5) to 23.6\% (N=10) within 3 months. 

\begin{figure}[h!]\label{strategy}
\caption{Time series plot of sentiment strategy investment return and investment return of $RMRF$. }
\centering
\includegraphics[width=0.75\textwidth]{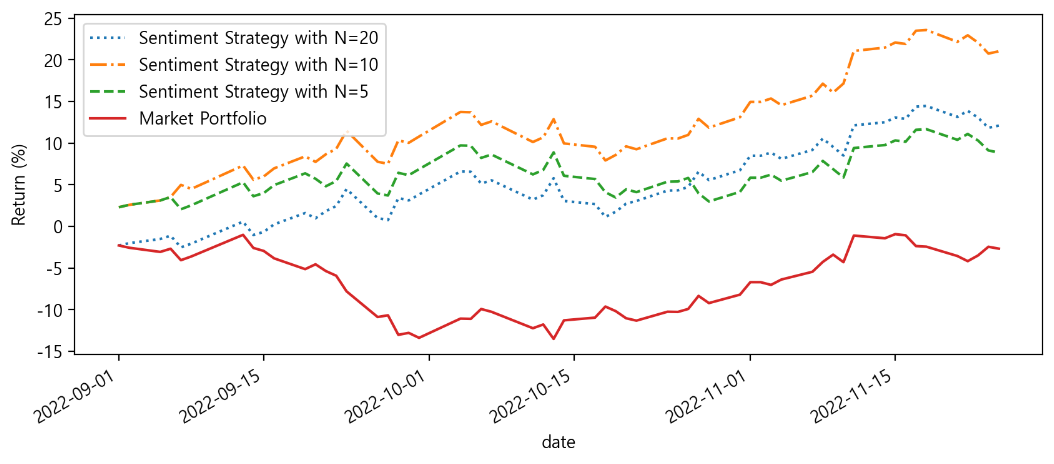}
\end{figure}

\subsection{Decomposing the Information Content on Market Timing}

Lastly, we report the Logit regression result to show our attempts to decompose the sentiment index's information content on market timing. The result is shown in Table \ref{regression}, which is little bit surprising. Even controlling lead variables of liquidity factor, interest rate and news sentiment change, $S(10, t-2)$ still show strong predictive power. We fail to decompose the sentiment index's predictive power in few specific ways. One part is that contradicting to theory of investor sentiment, $S(10, t-2)$ show positive correlation between future market ups and downs, unlike $\Delta NSI(t)$'s negative correlation with the market which is consistent with the theory. The other part is that $PctZero(t)$ and $R_f(t)$ failed to absorb $S(10, t-2)$'s predictive power, which could be due to the high-frequency nature of this analysis. We conduct the analysis with linear regression, and while we do not report the specifics, the result is fairly similar, with coefficient's statistical significance strengthen for the control variables.

\begin{table}[]
\centering
\begin{tabular}{ccccc}
\hline
\multicolumn{1}{l}{}                                             & $S(10, t-2)$& $\Delta NSI(t)$& $PctZero(t)$& $R_f(t)$\\ \hline
                                                                 & 0.52   &                                                 &                                                 &                                                \\
                                                                 & (2.02) &                                                 &                                                 &                                                \\
\multirow{-3}{*}{\textit{Logit Regression1}}                     &                                                  &                                                 &                                                 &                                                \\
                                                                 &                                                  & -3.11 &                                                 &                                                \\
                                                                 &                                                  & (-2.23)                                         &                                                 &                                                \\
\multirow{-3}{*}{\textit{Logit Regression2}}                     &                                                  &                                                 &                                                 &                                                \\
                                                                 &                                                  &                                                 & 27.55                                           & 0.66\\
                                                                 &                                                  &                                                 & (1.47)                                          & (0.85)                                         \\
\multirow{-3}{*}{\textit{Logit Regression3}}                     &                                                  &                                                 &                                                 &                                                \\
                                                                 & 0.60   & -3.45                                           &                                                 &                                                \\
                                                                 & (2.19)                                           & (-2.35)                                         &                                                 &                                                \\
\multirow{-3}{*}{\textit{Logit Regression4}}&                                                  &                                                 &                                                 &                                                \\
\multicolumn{1}{r}{}                                             & 0.77   & -4.00 & 46.04 & 0.96 \\
\multicolumn{1}{r}{\multirow{-2}{*}{\textit{Logit Regression5}}} & (2.48) & (-2.56)                                         & (1.97)                                          & (1.08)                                         \\ \hline
\end{tabular}
\caption{Five logistic regression result (regression slope values) that regress stock market's up(1) and down(0) by $S(10, t-2)$, $\Delta NSI(t)$ (the change of NSI, $PctZero(t)$, and short interest rate $R_f(t)$. Values in parentheses is the T-values that correspond to each slope above it. For each regression, we exclude or include some variables to compare the $S(10, t-2)$'s predictability when other variables are controlled.}
\label{regression}
\end{table}
Although we may not found the exact source of information content of $S(N, t)$, we can report on the following point. One is that, in this sample period, the $S(N, t)$, which was derived from 3PROTV, have some information content on market timing, and the information content is somewhat independent from news sentiment. We do not know $S(N, t)$ is a investor sentiment, however the regression result suggets otherwise. $S(N, t)$'s information content do not overlap with future $NSI$, which is the currently the best high frequency proxy for investor sentiment available for the public. The positive sign of the coefficent of $S(10, t-2)$ also show that the index is less likely to be a sentiment index, and rather some kind of fundamental information indicator. The other point is that the economic effect of $S(N, t)$ is quite strong and robust. Even for the worst parameter choice, it yield over 10\% excess return than the buy-and-hold market portfolio. However the tests for validity of  $S(N, t)$ is quite limited at this moment. With lack of long-term historcal data, currently we cannot concretely conclude for all time period that 3PROTV holds information content on market timing. 

However within our sample, we can carefully conclude that we found some potentials that 3PROTV could have information content on market timing.

\section{Conclusion}

Due to the lack of data, it is unwise to result in sweeping conclusion, but we discuss the result with a limited generality while focusing on its implication to investment strategy. First, we found that 3PROTV holds information content on stock selection, which means trading according to its information could enable outperformance. However, most of the outperformance is due to delayed reaction to negative information, and trading according to positive information will cause higher volatility and market timing disadvantages. Second, change in 3PROTV's sentimental tone have predictive power of future market return, implying trading according to 3PROTV does have some market timing advantage in the total market sense. 

It is interesting to point out that the content sentiment is more useful regarding overall market investment rather than stock selection. We suspect that this is because of the characteristics of general audience base of 3PROTV. As we discussed before, 3PROTV's main target customer is individual investor, not professionals or institutional investors. These individual investors are free from index-tracking obligations. Also it is unlikely for the audience of financial advisory YouTube channel to be satisfied with investment performance of index-tracking funds, thus the audience are more likely to incorporate the information about specific firms in the market, not about the total market. Based on this hypothesis, we conjecture that the individual stock-related information announced by 3PROTV is already incorporated into the price before the annoucement, but overall market-related information is not quickely incorporated into the price, leaving out the opportunity of arbitrage.

That being said, we also note that this result cannot be stretched to other areas, like generalizing about information content of all financial YouTube channels or even within 3PROTV's out-of-sample period. Also some might suggest that natrual language processing of the scripts were quite rough and loose (e.g. $SENTIMENT^{(i)}$ for stocks are more content sentiment rather than individual stock sentiment), making the result of stock selection analysis less reliable. Though we argue that this paper have contribution to the literature, like the test for information content on emerging market social media with the main contribution to the literature being not on stock selection analysis but on analyzing YouTube contents about their information contents on market timing, viewing financial YouTube contents in a time series perspective, whose value is not well recognised.

\vskip 0.2in

\bibliography{main}
\bibliographystyle{plain}
\end{document}